\documentclass{elsart3}
\usepackage{graphicx}
\begin{document}
\runauthor{Grzegorz Litak}

\begin{frontmatter}
\title{Specific Heat of Disordered Superconductors Induced
by Negative Centers}

\author{Grzegorz Litak\thanksref{E-mail}} 
\address{Department of Mechanics, Technical University of
Lublin,  Ul. Nadbystrzycka 36, PL--20-618  Lublin, Poland.}
\address{Max Planck Institute for Physics of Complex Systems,  
 N\"{o}thnitzer Str. 38, D--01187 Dresden, Germany. 
}

\thanks[E-mail]{Tel.: +48- 81- 5381573; Fax: +48- 81- 5241004; E-mail:
litak@archimedes.pol.lublin.pl}

\begin{abstract}
We show that superconductors with inhomogeneous  order parameters 
can show similar
features as anisotropic ones. 
In this paper we study
the low temperature specific heat dependence in  such a system 
and we show that the disorder associated with  randomly 
distributed attractive centers convert the BCS temperature exponential
 behaviour into a power law formula.   
To describe superconductivity we used a random version of the
 negative U Hubbard model, while the disorder was 
treated by means of the Coherent Potential Approximation (CPA).
 \end{abstract}
\end{frontmatter}

According to  the Anderson theorem the effect of non-magnetic disorder on 
s-wave superconductors can be neglected unless
spatial
fluctuations of the order parameter are negligible \cite{And59,Gyo96}. 
This theorem for 
conventional superconductors  
can also be  applied to other superconductors with  a long
coherence 
length where the 
fluctuations of
the order parameter induced by disorder 
\cite{Gyo96,Mor01}  are small and the quasiparticle density of states shows a clear gap.  As 
a consequence, superconducting properties, 
such as the specific heat and the magnetic penetration depth,  feature the 
typical
exponential BCS  dependences.
The situation is quite different for exotic superconductors. There the order parameter is modulated 
by 
a wave vector 
{\bf k} and even  disappears (for d-wave symmetry of the order parameter) 
at Fermi surface  line nodes. 
 
In the case of the recently discovered MgB$_2$ \cite{Nag01} 
with a high critical temperature $T_C \approx 40$ K and a relatively short coherence length $\xi$
(from 25 to 120\AA~,  depending on the sample \cite{Gab02}) the 
Anderson 
assumption of  a uniform order 
parameter
is also not satisfied.
In fact, there is a lot of evidence
\cite{Gab02} that
in this
compound the multi-gap scenario with a spatially fluctuating s-wave order parameter is 
realized.  Moreover Sharma et al. \cite{Sha03} found
a percolative transition of superconductivity in  Mg$_{1-x}$B$_2$. 

Prompted by these findings we examine the effects of a low
temperature specific heat dependence using the random 
negative U Hubbard model proposed to describe percolation in 
superconductors \cite{Lit00}.
We 
study the case where $U_i$ is $-|U|$ and 0 with probability $c$ and $1-c$,
respectively on a square lattice whose sites are labelled $i$ using the Gorkov
decoupling \cite{Mic90} and the Coherent Potential Approximation (CPA) \cite{Lit00}.

\begin{figure}[htb]
\begin{center}
~
\vspace{-0.6cm}

\includegraphics[angle=-90,width=13.0pc]{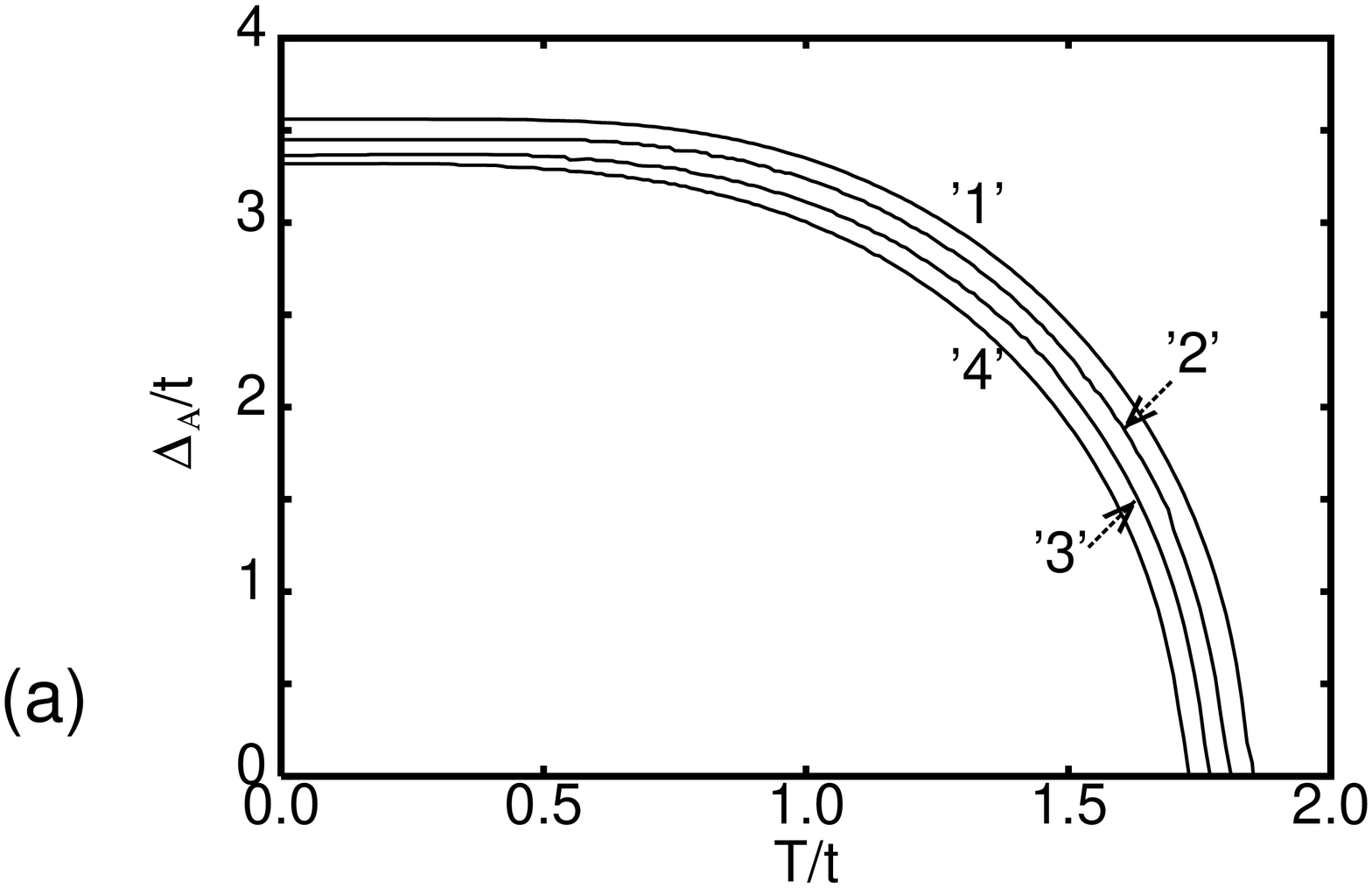}
\vspace{-0.6cm}

\includegraphics[angle=-90,width=13.0pc]{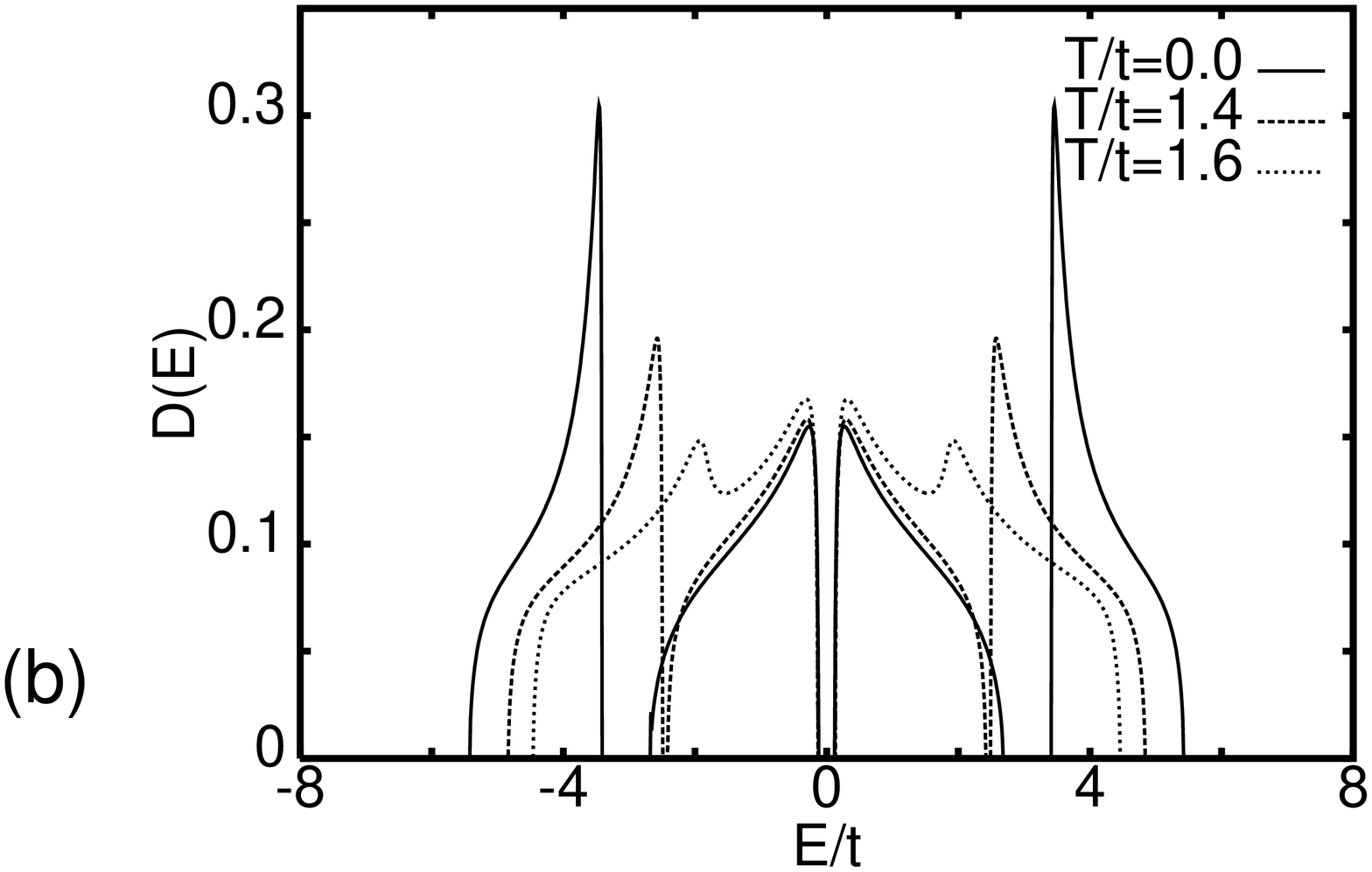}
\vspace{-0.6cm}

\includegraphics[angle=-90,width=12.8pc]{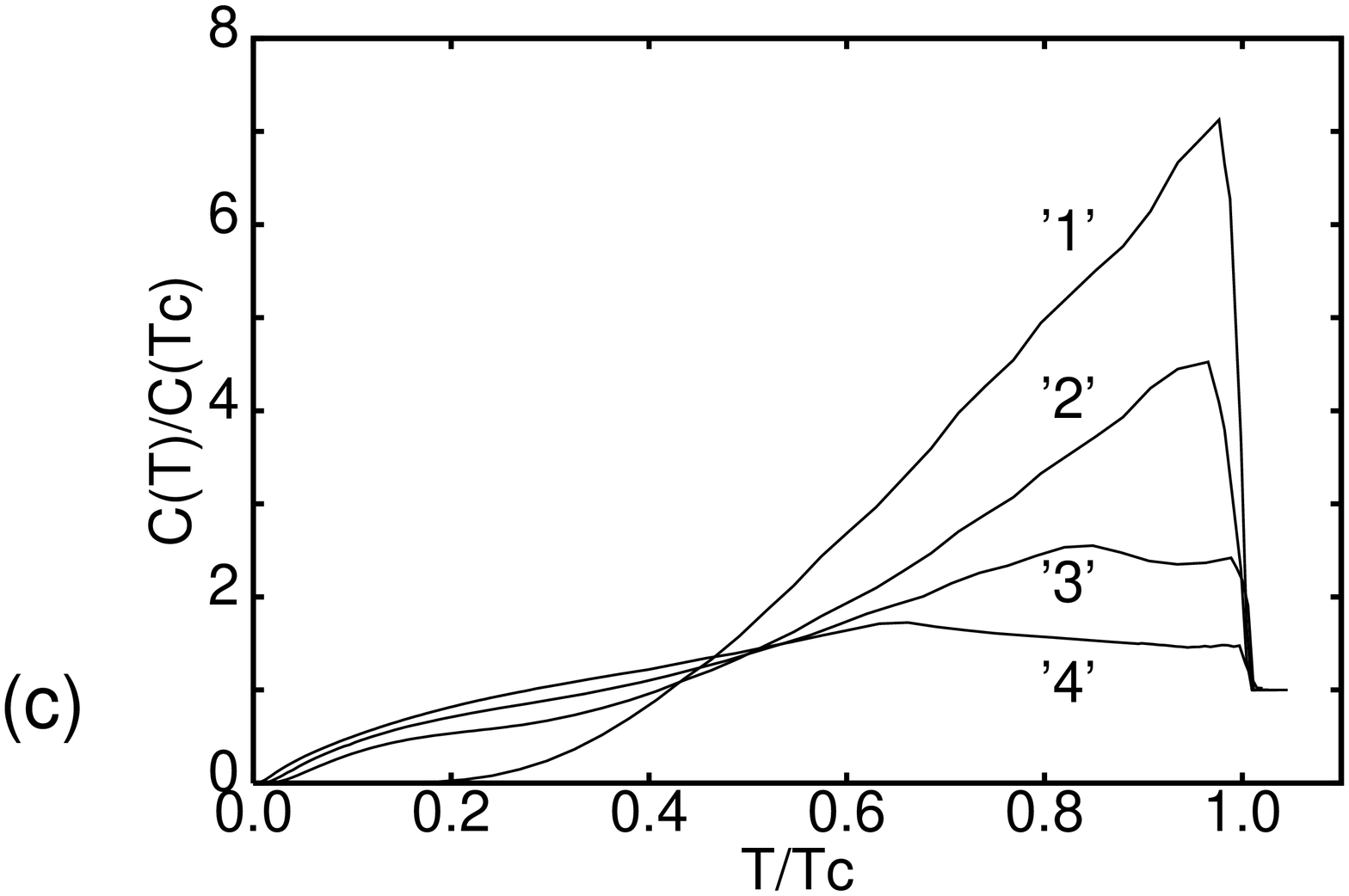}
\end{center}
\caption{ \label{fig:1}
Temperature dependence of $\Delta_A$ for various
 concentrations of negative centers $c$
in a binary alloy A$_c$B$_{1-c}$ (a).
Quasiparticle density of states for $c=0.5$ and three different 
temperatures (b).
Specific heat for superconductors with 
various concentration of negative centers (c).
In Figs. 1(a) and (c) lines denoted by '1--4'
correspond to  $c=0$, 0.7, 0.5, 0.3, respectively.}
\end{figure}

For the above assumptions we have done calculations of 
the order 
parameter $\Delta_A$ for various concentrations of negative centers $c$
in a binary alloy A$_c$B$_{1-c}$ \cite{Lit00}. Here, for simplicity, we neglected the Hartree term 
$Un/2$. 
This term may be important in some other contexts, as it is known to lead 
to a band splitting phenomenon 
and consequently to the appearance of the critical concentration of  a 
superconducting phase  $c_0$ 
\cite{Lit00}.


The temperature dependences of $\Delta_A$ for $U_A=-8t$ are presented in Fig. 
1a, while examples of qua\-siparticle densities of states (QDOS) for 
different temperatures are depicted 
in Fig. 1b. Note that 
the pairing potential 
  $\Delta_A$
for $c\neq 1$
 differs slightly from the uniform case (c=1), however the QDOS with states inside the gap is 
fundamentally 
different.    

Now the low temperature specific heat can be easily 
evaluated using the following formula \cite{Gab02,Gyo98}:  
\begin{equation}
 C=
2 T \int_{-\infty}^{\infty} D(\omega) \frac{\omega}{T} \frac{\rm d}{{\rm d} T}
\left(
\frac{1}{{\rm e}^{\omega /T} +1}  \right), 
\end{equation}
where $D(\omega)$ denotes QDOS.  

Fig. 1c shows the results for the specific heat for $c=1.0$, 0.7, 0.5 and 
0.3. 
Note that, in contrast
to a clean case ('1') where the corresponding 
plot can be associated with exponential formula $C/T 
\sim 
\exp(-\alpha/T)$
for constant $\alpha$, 
the next curves ('2-4') mimic a distinct power law behaviour (excluding 
some lowest temperatures).

Thus, the behaviour of  superconductors with randomly distributed negative centers showing 
fluctuations of the
pairing
potential in real space resembles, in some sense, the behaviour of 
anisotropic
but clean systems with a pairing potential varying in {\bf k} space. 
Namely,   a wave vector 
modulation of the superconducting 
order 
parameter in the case 
of  s-wave and most of all line nodes in the case of d-wave 
superconductors lead to  similar 
effects with a  low temperature  unusual
dependence of the specific heat.

This work has been partially supported by the KBN  Grant
No. 3P03B06225.

\end{document}